\documentclass[aps,prd,epsecnum,preprint,nofootinbib]{revtex4}
\usepackage{bm}
\usepackage{graphicx}
\usepackage{color}
\usepackage{amsmath,amsthm,amssymb}

\begin{document}
\baselineskip5.5mm
\title{
Comparison of two approximation schemes for solving perturbations in a LTB cosmological model
}
\author{
       ${}^{1}$Ryusuke Nishikawa \footnote{E-mail:ryusuke@sci.osaka-cu.ac.jp},
	${}^{1}$Ken-ichi Nakao \footnote{E-mail:knakao@sci.osaka-cu.ac.jp},
and
       ${}^{2}$Chul-Moon Yoo \footnote{E-mail:yoo@gravity.phys.nagoya-u.ac.jp},
}
\affiliation{
${}^{1}$Department of Mathematics and Physics,
Graduate School of Science, Osaka City University,
3-3-138 Sugimoto, Sumiyoshi, Osaka 558-8585, Japan
\\
${}^{2}$Division of Particle and Astrophysical Science, 
Graduate School of Science, Nagoya University, 
Furo-cho, Chikusa-ku, Nagoya 464-8602, Japan
}

\begin{abstract}
\baselineskip5.5mm
Recently, the present authors studied perturbations
in the Lema\^{\i}tre-Tolman-Bondi cosmological model by applying the second-order 
perturbation theory in the dust Friedmann-Lema\^{\i}tre-Robertson-Walker universe model.
Before this work, the same subject was studied in some papers by analyzing linear 
perturbations in the Lema\^{\i}tre-Tolman-Bondi cosmological model under 
the assumption proposed by Clarkson, Clifton and February, in which two of perturbation variables are negligible. 
However, it is a non-trivial issue in what situation the Clarkson-Clifton-February assumption is valid.  
In this paper, 
we investigate differences between these two approaches. 
It is shown that, in general, these two approaches are not compatible with 
each other. 
That is, in our 
perturbative procedure, the Clarkson-Clifton-February assumption is 
not valid at the order of our interest. 
\end{abstract}

\preprint{OCU-PHYS 404}

\preprint{AP-GR 111}

\maketitle

\section{introduction}
Most of modern cosmological models are based on the Copernican principle 
which states we are not living at a privileged position in the universe. 
The observed isotropy of the Cosmic Microwave Background (CMB) radiation 
together with the Copernican principle 
implies our universe is homogeneous and isotropic,
if the small scale structures less than 50 Mpc are coarse-grained.
Although the standard cosmology can explain a lot of observations naturally, 
it should be noted that the Copernican principle on cosmological scales $\geq$ 1 Gpc has not been confirmed. 
This means modern cosmology would contain systematic errors
that arise from non-Copernican inhomogeneities of the background universe, which is usually 
unexpected. The systematic errors may mislead us 
when we consider major issues in modern cosmology such as testing general relativity at cosmological scales and probing dark energy abundance. 
Thus, it is an essential task in modern precision cosmology to test the Copernican principle. 

In order to test the Copernican principle, 
we have to investigate non-Copernican cosmological models which drop the Copernican principle. 
Non-Copernican models commonly assume that we live close to the center in a spherically 
symmetric spacetime since the universe is observed to be nearly isotropic around us. 
These models have also been studied as an alternative to dark energy, 
because some of them can explain the observational data of the luminosity distances of Type Ia supernovae without introducing dark energy~\cite{Bull:2012zx,Celerier:1999hp,Celerier:2009sv,Clifton:2008hv,Goodwin:1999ej,Iguchi:2001sq,Kolb:2009hn,Mustapha:1998jb,Tomita:1999qn,Tomita:2000jj,Tomita:2001gh,Vanderveld:2006rb,Yoo:2008su,Yoo:2010qn}.
The non-Copernican models without dark energy have been tested by observations including the CMB acoustic peaks~\cite{Alexander:2007xx,Alnes:2005rw,Biswas:2010xm,Bolejko:2008cm,Clarkson:2010ej,GarciaBellido:2008nz,Marra:2011ct,Marra:2010pg,Moss:2010jx,Nadathur:2010zm,Yoo:2010qy,Zibin:2008vk},
the present Hubble parameter $H_0$~\cite{Biswas:2010xm,Bolejko:2008cm,GarciaBellido:2008nz,Marra:2011ct,Moss:2010jx},
the Baryon Acoustic Oscillation scale of the galaxy correlations~\cite{Biswas:2010xm,GarciaBellido:2008yq,Zumalacarregui:2012pq},
the kinematic Sunyaev-Zeldovich effect~\cite{Bull:2011wi,GarciaBellido:2008gd,Moss:2011ze,Yoo:2010ad,Zhang:2010fa,Ade:2013opi}
and others~\cite{Adachi:2011vu,Alnes:2006uk,Bolejko:2011jc,Bolejko:2005fp,Caldwell:2013fua,Celerier:2012xr,Clarkson:2012bg,dePutter:2012zx,Dunsby:2010ts,Enqvist:2009hn,Enqvist:2006cg,Goto:2011ru,Heavens:2011mr,Quartin:2009xr,Regis:2010iq,Romano:2009mr,Romano:2010nc,Romano:2011mx,Tanimoto:2009mz,Tomita:2009ar,Yagi:2012vx,Zibin:2011ma},
and significant observational constraints exist, although they have never been completely excluded yet.
However, it should be noted that even if we accept dark energy components, 
the existence of the large spherical inhomogeneity may significantly affects observational results
(see e.g. Ref.~\cite{Valkenburg:2013qwa,Valkenburg:2012td}).
A large void universe model which assumes we live at a center of a huge spherical void  
whose radius is larger than 1 Gpc is known as one of popular models among the non-Copernican cosmologies,
and we take such model into consideration in this paper. 

Growth of the large-scale structure in the universe can be thought of 
as one of the most useful tools to investigate the large spherical void model,
because the evolution of perturbations is expected to reflect the tidal force in the background spacetime.
Unfortunately, linear perturbation equations in the spherical void universe 
cannot be solved analytically in general situation~\cite{Gerlach:1980},
because the number of isometries in a spherically symmetric inhomogeneous spacetime  is too little. 

Recently, linear perturbations in the LTB cosmological model and the observations 
related to them have been studied by several researchers\cite{Dunsby:2010ts,Alonso:2012ds,Alonso:2010zv,Clarkson:2009sc,February:2012fp,February:2013qza,Zibin:2008vj}:  
Zibin~\cite{Zibin:2008vj} and Dunsby, Goheer, Osano and Uzan~\cite{Dunsby:2010ts} solved perturbations by using 
``silent universe approximation'' in which the magnetic part of the Weyl tensor vanishes.  
However, we should note that the magnetic part of the Weyl tensor usually plays an important role even in Newtonian situations~\cite{Bertschinger:1994nc}, and hence the silent universe model may fail to 
include the situations of our interest. 
Clarkson Clifton and February~\cite{Clarkson:2009sc} (hereafter CCF) classified perturbations into the ``scalar'', ``vector'' and ``tensor'' degrees of freedom by 
their behaviors in the limit to the homogeneous and isotropic background, 
and solved perturbation equations by assuming that the ``vector'' and ``tensor'' modes are negligible. 
However, in the case of the LTB spacetime, the vector and tensor modes couple to the scalar mode in general.
It is not clear whether CCF's assumption is valid in the situations of our interest. 
In our previous works~\cite{Nishikawa:2012we,Nishikawa:2013rna}, 
we studied the same subject by analyzing perturbations in the Einstein-de Sitter(EdS) universe model 
up to the second order without any additional assumptions 
and revealed the growth rate of the density perturbations and their two point 
correlation functions. In this paper,  
we study the validity of  CCF's assumption 
within the non-linear perturbation approach that we have adopted 
in our previous work{\bf s}. 

We adopt the same sign conventions of the metric and the Riemann tensors as those in 
Ref.~\cite{HE}.

\section{A review on previous work by Clarkson-Clifton-February}

First, we review linear perturbations in the LTB spacetime.
The infinitesimal world interval and the stress-energy tensor are given by 
\begin{eqnarray}
 ds^2&=&-dt^2+\frac{a_{||}^2(t,r)}{1-k(r)r^2}dr^2+a_\bot^2(t,r)r^2d\Omega^2, 
 \label{C:LTBds} \\
 T_{\mu\nu}&=&\rho^{\rm LTB}(t,r) \bar{u}_\mu \bar{u}_\nu,
 \label{C:LTBTmn}
\end{eqnarray}
where $a_{||}=\partial_r (r a_\bot)$ and $\bar{u}_\mu =(-1,0,0,0)$.
Because of the spherical symmetry  of the LTB spacetime, perturbations in the LTB 
spacetime can be decoupled into two independent modes, 
called the polar and the axial modes (for details, see~\cite{Clarkson:2009sc}).
Since we are interested in the evolution of the density perturbations, we focus on the polar mode. 
The metric of the perturbed LTB spacetime in the Regge-Wheeler (RW) gauge is written as
\begin{eqnarray}
 ds^2&=&-\left[1+(2\tilde{\eta}-\tilde{\chi}-\tilde{\varphi})\right]dt^2
 -2\frac{\tilde{\varsigma}a_{||}}{\sqrt{1-kr^2}}dtdr \cr
 &+&\left[1+(\tilde{\chi}+\tilde{\varphi})\right]\frac{a_{||}^2}{1-kr^2}dr^2
 +a_\bot^2r^2(1+\tilde{\varphi})d\Omega^2,
 \label{C:linds}
\end{eqnarray}
where $\tilde{\eta}(t,{\bf x})$, $\tilde{\chi}(t,{\bf x})$, $\tilde{\varphi} (t,{\bf x})$ and 
$\tilde{\varsigma} (t,{\bf x})$ are polar perturbations,
and their expansion coefficients of a spherical harmonic expansion
correspond to $\eta(t,r)$, $\chi(t,r)$, $\varphi(t,r)$ and $\varsigma(t,r)$ in 
CCF's paper, respectively~\cite{Clarkson:2009sc}.
The density and 4-velocity of the perturbed spacetime are given by
\begin{eqnarray}
 \rho&=&\rho^{\rm LTB}\left(1+\tilde{\delta}\right), 
 \label{C:lindel} \\
 & &\cr
 u_\mu &=&
 \left(-1-\frac{1}{2}(2\tilde{\eta}-\tilde{\chi}-\tilde{\varphi}),
 \frac{a_{||}\left(\tilde{w}-\tilde{\varsigma}/2\right)}{\sqrt{1-kr^2}},\partial_\theta \tilde{v},\partial_\varphi \tilde{v}\right),
 \label{C:linumu}
\end{eqnarray}
where $\tilde{\delta}(t,{\bf x})$, $\tilde{w}(t,{\bf x})$ and $\tilde{v}(t,{\bf x})$ are polar perturbations,
and their expansion coefficients of a spherical harmonic expansion
correspond to $\Delta(t,r)$, $w(t,r)$ and $v(t,r)$ in CCF's paper, respectively. 
By substituting the expressions~(\ref{C:linds}), (\ref{C:lindel}) and (\ref{C:linumu}) into the Einstein equations,
we obtain the perturbation equations (see Eqs.~(3.1)--(3.7) in CCF). From one of the perturbation equations, 
we obtain $\tilde{\eta}=0$ in general (see Eq.~(3.4) in CCF).
However, other perturbation equations are 
complicated and cannot be reduced to ordinary differential 
equations differently from
that in the homogeneous and isotropic universe models.

In section 3 of Ref.~\cite{Clarkson:2009sc}, CCF analyzed the equation for 
the density perturbation by assuming
\begin{eqnarray}
 \tilde{\chi}=\tilde{\varsigma}=0.
 \label{C:CCF0}
\end{eqnarray}
By virtue of this assumption, 
the perturbation equations are reduced to very simple forms (see also~\cite{February:2012fp}), and hence 
CCF succeeded in obtaining the analytic solutions of  the equations for the perturbations, $\tilde{\varphi}$ 
and $\tilde{\delta}$. 
However, the range of the validity of this assumption is not clear in
contrast to our perturbative procedure. 
We check the validity within our perturbative procedure in the next section.

\section{Perturbative analysis of CCF procedure}

In our previous works~\cite{Nishikawa:2012we,Nishikawa:2013rna}, we studied perturbations in 
the LTB universe model by applying the second-order perturbation theory 
for the dust-filled FLRW universe model. We introduced perturbations 
parametrized by two small expansion parameters $\kappa$ and $\epsilon$. 
The limit $\epsilon\rightarrow0$ leads to the exact 
LTB solution, if we take all orders of $\kappa$ into account.
By contrast, the limit 
$\kappa\rightarrow0$ with $0<\epsilon\ll1$ leads to the homogeneous and 
isotropic universe with small anisotropic perturbations. 
Then, in order to see the effect of the non-Copernican structure 
on the evolution of the anisotropic perturbations, we studied the non-linear 
effects up to the order of $\kappa\epsilon$.
Here, it should be noted that we included only the scalar mode for perturbations of 
the order $\epsilon$, since the vector and the tensor modes of the linear perturbations 
in the homogeneous and isotropic universes do not grow with time as is well known.
We studied perturbations in the synchronous comoving gauge,
while CCF studied in the RW gauge.
To compare the two approaches, we derive perturbation equations of our approach in the RW gauge.

In order to apply our method, we regard the background LTB spacetime in 
Eqs.~(\ref{C:LTBds}) and (\ref{C:LTBTmn}) as isotropic perturbations 
in the EdS universe model which are characterized by 
$\kappa$ which should be regarded as a book-keeping parameter here: 
it is replaced by unity after deriving the perturbation equations. 
By introducing isotropic perturbations
$\ell_{||}^{(1)}(t,r)$, $\ell_\bot^{(1)}(t,r)$ and $\Delta^{(1)}(t,r)$ (for details, see~\cite{Nishikawa:2012we}),
we write the functions $a_{||}$, $a_\bot$, $k(r)$ and $\rho^{\rm LTB}$  in the form
\begin{eqnarray}
 a_{||}&=&a\left[1+\frac{\kappa}{2}\partial_r(r\ell^{(1)}_\bot)\right]+\mathcal{O}(\kappa^2), 
 \label{3:exp1}
 \\
 a_\bot &=&a\left[1+\frac{\kappa}{2}\ell^{(1)}_\bot \right]+\mathcal{O}(\kappa^2), 
 \label{3:exp2} \\
 k(r)r^2&=&\kappa\left[\ell^{(1)}_{||}-\partial_r(r\ell^{(1)}_\bot)\right]+\mathcal{O}(\kappa^2),
 \label{3:exp3} \\
 \rho^{\rm LTB}&=&\bar{\rho}\left[1+\kappa\Delta^{(1)}\right]+\mathcal{O}(\kappa^2),
 \label{3:exp4}
\end{eqnarray}
where $a$ and $\bar{\rho}$ are the scale factor and the matter density in the background EdS universe.
In accordance with the prescription in our previous paper~\cite{Nishikawa:2012we}, we write the polar modes of perturbations as follows:
\begin{eqnarray}
 \tilde{\varphi}
 =\epsilon \tilde{\varphi}^{(1)} +\kappa \epsilon \tilde{\varphi}^{(2)}+\mathcal{O}(\kappa^2\epsilon),
 \label{3:exp5}
\end{eqnarray}
and $\tilde{\chi}$, $\tilde{\varsigma}$, $\tilde{\delta}$, $\tilde{w}$ and $\tilde{v}$ are written 
in the similar forms.
Assuming $\epsilon\ll \kappa <1$ and the Einstein equations hold in each order with respect to $\kappa$ and $\epsilon$,
we can obtain the equations for the perturbations of the order $\epsilon$ and $\kappa\epsilon$.

First of all, we see the perturbation equations of the order $\epsilon$.
The perturbations of the order $\epsilon$ are 
equivalent to those of the EdS universe model.
As mentioned above, we only include the scalar mode of the order $\epsilon$ in our present analysis.
By contrast, the polar perturbations 
may be a mixture of the scalar, vector and tensor modes 
in the context of the perturbation theory for the EdS universe model.
Thus, we need to relate the polar perturbations to the scalar, vector and tensor
perturbations in the EdS universe model. 
CCF derived these relations in Eqs.~(4.5)--(4.8) in their paper, 
by taking  the homogeneous limit of LTB universe model. 
From Eqs.~(4.6) and (4.7) in CCF,
we can see that $\tilde{\varsigma}^{(1)}$ and $\tilde{\chi}^{(1)}$ 
do not contain the scalar mode but the vector and tensor modes. 
Thus, the neglect of the vector and tensor modes leads to
\begin{eqnarray}
 \tilde{\varsigma}^{(1)}=\tilde{\chi}^{(1)}&=&0.
 \label{3:RW1}
\end{eqnarray}
In the case that Eq.~(\ref{3:RW1}) holds, 
up to the order of $\epsilon$,
the perturbed metric (\ref{C:linds}) becomes 
\begin{equation}
ds^2=-\left(1-\tilde{\varphi}^{(1)}\right)dt^2+a^2\left(1+\tilde{\varphi}^{(1)}\right)\left(dr^2+r^2d\Omega^2\right).
\end{equation}
It is clear from the above equation that the metric perturbation of the order $\epsilon$ is the scalar mode. 
Correspondingly, the perturbations of $\rho$ and $u^\mu$ of the order $\epsilon$ contain the only scalar mode.

By substituting the expressions~(\ref{C:linds}), (\ref{C:lindel}) and (\ref{C:linumu}) into the Einstein equations
and using the expansions~(\ref{3:exp1})-(\ref{3:exp5}) 
together with Eq.~(\ref{3:RW1}), we obtain the perturbation equations of the order $\epsilon$ as
\begin{eqnarray}
 \ddot{\tilde{\varphi}}^{(1)}+4H\dot{\tilde{\varphi}}^{(1)}
 &=&0, 
 \label{3:first1} \\
 8\pi\bar{\rho}\tilde{\delta}^{(1)}
 &=&
 3H\dot{\tilde{\varphi}}^{(1)}+3H^2\tilde{\varphi}^{(1)}
 -\frac{1}{a^2}\mathcal{D}^i\mathcal{D}_i\tilde{\varphi}^{(1)}, 
 \label{3:first2} \\
 8\pi\bar{\rho}\tilde{w}^{(1)}
 &=&
 \frac{\partial_r}{a}\left(\dot{\tilde{\varphi}}^{(1)}+H\tilde{\varphi}^{(1)}\right), 
 \label{3:first3} \\
 8\pi\bar{\rho}\tilde{v}^{(1)}
 &=&
 \dot{\tilde{\varphi}}^{(1)}+H\tilde{\varphi}^{(1)},
 \label{3:first4} 
\end{eqnarray}
where we defined the operator $\mathcal{D}^i\mathcal{D}_i$ as
\begin{eqnarray}
 \mathcal{D}^i\mathcal{D}_i:=
 \frac{1}{r^2}\frac{\partial}{\partial r}\left(r^2\frac{\partial}{\partial r}\right)
 +\frac{1}{r^2\sin\theta}\frac{\partial}{\partial \theta}\left(\sin\theta \frac{\partial}{\partial\theta}\right)
 +\frac{1}{r^2\sin^2\theta}\frac{\partial^2}{\partial^2\phi}.
\end{eqnarray}
Although it is not clear from only Eq.~(\ref{C:linumu}) 
whether the perturbations of $u_\mu$ of the order $\epsilon$  
are the scalar mode or the mixture 
with the vector mode, 
it is clear from Eqs.~(\ref{3:first3}) and (\ref{3:first4}) that it is 
purely a scalar mode.
Thus, the assumption (\ref{C:CCF0}) up to the order $\epsilon$ 
is equivalent to neglecting the vector and tensor modes. 

Next, we see the perturbation equations of the order $\kappa\epsilon$.
Substituting the expressions~(\ref{C:linds})--(\ref{C:linumu}) 
into the Einstein equations and using the expansions~(\ref{3:exp1})--(\ref{3:exp5}),
we obtain the perturbation equations of the order $\kappa\epsilon$ as
\begin{eqnarray}
 & &
 -\ddot{\tilde{\chi}}^{(2)}-3H\dot{\tilde{\chi}}^{(2)}+\frac{1}{a^2}
 \left[\mathcal{D}^i\mathcal{D}_i-2\left(\frac{2}{r}\partial_r-\frac{1}{r^2}\right)\right]
 \tilde{\chi}^{(2)}
 =
 2(\dot{\ell}^{(1)}_{||}-\dot{\ell}^{(1)}_\bot)\dot{\tilde{\varphi}}^{(1)} \cr
&&\cr
 & &
 ~~+2\left[\left(\ddot{\ell}^{(1)}_{||}-\ddot{\ell}_\bot^{(1)}\right)
 +3H\left(\dot{\ell}^{(1)}_{||}-\dot{\ell}^{(1)}_\bot\right)\right]\tilde{\varphi}^{(1)},
 \label{3:second1} \\
&&\cr
& &\ddot{\tilde{\varphi}}^{(2)}+4H\dot{\tilde{\varphi}}^{(2)}
 +H\dot{\tilde{\chi}}^{(2)}+\frac{1}{2a^2}
 \left[\mathcal{D}^i\mathcal{D}_i-\partial_r^2-2\left(\frac{2}{r}\partial_r-\frac{1}{r^2}\right)
 \right]\tilde{\chi}^{(2)}
 =-2\dot{\ell}_\bot^{(1)}\dot{\tilde{\varphi}}^{(1)} \cr
&&\cr
& &~~+\frac{2}{a^2r^2}
 \left(\ell_{||}^{(1)}-\partial_r (r\ell_\bot^{(1)})\right)\tilde{\varphi}^{(1)},
 \label{3:second2} \\
&&\cr
 & &\dot{\tilde{\varsigma}}^{(2)}+2H\tilde{\varsigma}^{(2)}+\frac{\partial_r\tilde{\chi}^{(2)}}{a}=0,
 \label{3:second3}
\end{eqnarray}

\begin{eqnarray}
 8\pi \bar{\rho}\tilde{\delta}^{(2)}
 &=&
 3H\dot{\tilde{\varphi}}^{(2)}+3H^2\tilde{\varphi}^{(2)}
 -\frac{1}{a^2}\mathcal{D}^i\mathcal{D}_i\tilde{\varphi}^{(2)} \cr
 &&\cr
&+&
 H\dot{\tilde{\chi}}^{(2)}+\left[3H^2+\frac{1}{2a^2}
 \left(\partial^2_r+\frac{4}{r}\partial_r+\frac{2}{r^2}-{\cal D}^i{\cal D}_i\right)\right]\tilde{\chi}^{(2)}
 +\frac{2H}{a}\left[\partial_r+\frac{2}{r}\right]\tilde{\varsigma}^{(2)} \cr
&&\cr
 &+&
 \frac{1}{a^2}\left(\ell_{||}^{(1)}-\ell_\bot^{(1)}\right)\partial_r^2\tilde{\varphi}^{(1)}
 +\frac{1}{a^2}\left[\frac{2}{r}\left(\ell_{||}^{(1)}-\ell_\bot^{(1)}\right)
 -\frac{1}{2}\partial_r\left(\ell_{||}^{(1)}-2\ell_\bot^{(1)}\right)\right]
 \partial_r\tilde{\varphi}^{(1)} \cr
&&\cr
 &+&\frac{1}{a^2}\ell^{(1)}_\bot \mathcal{D}^i\mathcal{D}_i\tilde{\varphi}^{(1)}
 +\frac{1}{2}\left(\dot{\ell}_{||}^{(1)}+2\dot{\ell}_\bot^{(1)}\right)\dot{\tilde{\varphi}}^{(1)}
 \cr
&&\cr
 &+&\left[2H\left(\dot{\ell}_{||}^{(1)}+2\dot{\ell}_\bot^{(1)}\right)
 -8\pi\bar{\rho}\Delta^{(1)}\right]\tilde{\varphi}^{(1)}
 -8\pi\bar{\rho}\Delta^{(1)}\tilde{\delta}^{(1)},
 \label{3:second4}
\end{eqnarray}
\begin{eqnarray}
 8\pi\bar{\rho}\tilde{w}^{(2)}
 &=&
 \frac{\partial_r}{a}\left(\dot{\tilde{\varphi}}^{(2)}+H\tilde{\varphi}^{(2)}\right)
 -\frac{1}{ar}\dot{\tilde{\chi}}^{(2)}+\frac{H}{a}\partial_r \tilde{\chi}^{(2)}
 +\frac{1}{2a^2}\left[\partial_r^2+\frac{2}{r}\partial_r-\mathcal{D}^i\mathcal{D}_i\right]\tilde{\varsigma}^{(2)}
 \cr
&&\cr
 &+&
 \frac{3}{2}H^2\tilde{\varsigma}^{(2)}
 -\frac{1}{2a}\ell_{||}^{(1)}\partial_r\dot{\tilde{\varphi}}^{(1)}
 -\frac{1}{2a}\left[H\ell_{||}^{(1)}-(2\dot{\ell}_\bot^{(1)}-\dot{\ell}_{||}^{(1)})\right]
 \partial_r\tilde{\varphi}^{(1)} \cr
&&\cr
 &-&8\pi\bar{\rho}\Delta^{(1)}\tilde{w}^{(1)},
 \label{3:second5}
\end{eqnarray}
\begin{eqnarray}
 8\pi\bar{\rho}\tilde{v}^{(2)}
 &=&
 \dot{\tilde{\varphi}}^{(2)}+H\tilde{\varphi}^{(2)}+\frac{1}{2}\dot{\tilde{\chi}}^{(2)}+H\tilde{\chi}^{(2)}
 +\frac{1}{2a}\partial_r \tilde{\varsigma}^{(2)}
 +\frac{1}{2}\dot{\ell}_{||}^{(1)}\tilde{\varphi}^{(1)}
 -8\pi\bar{\rho}\Delta^{(1)}\tilde{v}^{(1)}.
 \label{3:second6}
\end{eqnarray}
By introducing the new variable $\tilde{y}$ defined as 
\begin{equation}
\tilde{y}:=\frac{\tilde{\chi}^{(2)}}{r^2},
\label{y:def1}
\end{equation}
and defining its Fourier transformation $\tilde{y}_{\bf k}$ as 
\begin{eqnarray}
 \tilde{y}(t,{\bf x})=\int \frac{d^3k}{(2\pi)^{3/2}}e^{i{\bf k\cdot x}}\tilde{y}_{\bf k}(t),
\end{eqnarray}
equation~(\ref{3:second1}) is reduced to 
\begin{eqnarray}
 \ddot{\tilde{y}}_{\bf k}+3H\dot{\tilde{y}}_{\bf k}+\frac{k^2}{a^2}\tilde{y}_{\bf k}
 &=&S_{\bf k}(t),
 \label{3:y1}
\end{eqnarray}
where the source term $S_{\bf k}(t)$ is defined as
\begin{eqnarray}
 S_{\bf k}(t)&:=&-\int \frac{d^3x}{(2\pi)^{3/2}}e^{-i{\bf k\cdot x}}\Bigg\{
 2(\dot{\ell}^{(1)}_{||}-\dot{\ell}^{(1)}_\bot)\frac{\dot{\tilde{\varphi}}^{(1)}}{r^2}
 +2\Big[\left(\ddot{\ell}^{(1)}_{||}-\ddot{\ell}_\bot^{(1)}\right) \cr
 & &
 +3H\left(\dot{\ell}^{(1)}_{||}-\dot{\ell}^{(1)}_\bot\right)\Big]
 \frac{\tilde{\varphi}^{(1)}}{r^2}
 \Bigg\}.
\end{eqnarray}
We find from Eqs.~(6)--(8) in Ref.~\cite{Nishikawa:2012we} that 
$\ell_{||}^{(1)}$ and $\ell_\bot^{(1)}$ are proportional to the scale factor $a$, 
and the growing solution of Eq.~(\ref{3:first1}) is $\tilde{\varphi}^{(1)}=$ constant. 
Hence, the source term $S_{\bf k}(t)$ is proportional to $a^{-2}$.
From this fact, we can see that the particular solution of Eq.~(\ref{3:y1}) is $S_{\bf k}(t_0)/k^2$, 
where we set $a(t_0)=1$ by using the freedom of the constant re-scaling in the spatial coordinates. 
By using the fact that the Hubble function 
and the scale factor are given as $H=2/(3t)$ and $a=(t/t_0)^{2/3}$ in the EdS universe model, 
we obtain the homogeneous solutions of Eq.~(\ref{3:y1}) as
\begin{eqnarray}
 \tilde{y}_{{\bf k},1}(t)=\frac{1}{kt_0^{2/3}t^{1/3}}j_1(3kt_0^{2/3}t^{1/3}),~~~~{\rm and}~~~~
 \tilde{y}_{{\bf k},2}(t)=\frac{1}{kt_0^{2/3}t^{1/3}}y_1(3kt_0^{2/3}t^{1/3}),
 \label{homo1}
\end{eqnarray}
where $j_1$ and $y_1$ are spherical Bessel function of the first kind and that of the second kind, respectively.
By using the particular solution and the homogeneous solutions~\eqref{homo1},
we obtain the general solution of Eq.~(\ref{3:y1}) as 
\begin{eqnarray}
 \tilde{y}_{\bf k}(t)=\frac{S_{\bf k}(t_0)}{k^2}
 +C_1({\bf k})\tilde{y}_{{\bf k},1}(t)+C_2({\bf k})\tilde{y}_{{\bf k},2}(t),
 \label{y:solu1}
\end{eqnarray}
where $C_1$ and $C_2$ are arbitrary functions.
From Eqs.~\eqref{y:def1} and \eqref{y:solu1}, 
we obtain the metric perturbation $\tilde{\chi}^{(2)}$ as 
\begin{eqnarray}
 \tilde{\chi}^{(2)}(t,{\bf x})=r^2\int \frac{d^3k}{(2\pi)^{3/2}}e^{i{\bf k\cdot x}}\frac{S_{\bf k}(t_0)}{k^2}
 +({\rm decaying~modes}).
 \label{3:chi1}
\end{eqnarray}
By using Eqs.~\eqref{3:second3} and \eqref{3:chi1}, we obtain
\begin{eqnarray}
 \tilde{\varsigma}^{(2)}(t,{\bf x})=-\frac{3}{5}t^{1/3}t_0^{2/3}\partial_r\tilde{\chi}^{(2)}(t,{\bf x})
 +({\rm decaying~modes}).
 \label{3:varsigma1}
\end{eqnarray}
From Eqs.~\eqref{3:chi1} and \eqref{3:varsigma1}, 
we can see that $\tilde{\chi}^{(2)}$ and $\tilde{\varsigma}^{(2)}$
do not vanish: In general, $\tilde{\chi}^{(2)}$ is temporally constant and $\tilde{\varsigma}^{(2)}$ is proportional to 
$t^{1/3}$ at late time, even if we set $\tilde{\varsigma}^{(1)}=\tilde{\chi}^{(1)}=0$, or in other words, 
$\tilde{\varsigma}$ and $\tilde{\chi}$ vanish up to the order $\epsilon$ . 
Thus, exactly speaking, the assumption proposed by CCF, i.e., Eq.~(\ref{C:CCF0}) is not valid, 
if we take the order $\kappa\epsilon$ into account in the evolution of metric perturbations.  
However, it is worthwhile to notice that the effects of $\tilde{\chi}^{(2)}$ and $\tilde{\varsigma}^{(2)}$ to
the density perturbation $\tilde{\delta}^{(2)}$ are less dominant than the other effects,  
because the terms with respect to $\tilde{\chi}^{(2)}$ and $\tilde{\varsigma}^{(2)}$ 
in the right hand side of Eq.~(\ref{3:second4}) proportional to $t^{-4/3}$ 
whereas the terms with respect to $\tilde{\varphi}^{(1)}$ proportional to $t^{-2/3}$.
Thus, the assumption proposed by CCF 
will be a good approximation for sufficiently late time in the study of the density perturbations.

We expect that $\tilde{\chi}^{(2)}$ and $\tilde{\varsigma}^{(2)}$ become important in testing LTB cosmological models
from observations of large-scale structures such as weak gravitational lensing,
since a photon from a galaxy propagates in the spacetime with these metric perturbations.
We leave the analysis of the effects to the observations for a future work.

February, Larena, Clarkson and Pollney (hereafter FLCP)~\cite{February:2013qza} 
numerically studied the perturbations in the LTB spacetime without any approximations except for 
the linearization and discretization of the differential equations governing them,
and investigated the validity of the assumption proposed by CCF.
FLCP showed that if the coupling of $\tilde{\delta}$ and $\tilde{\varphi}$ 
to $\tilde{\chi}$ and $\tilde{\varsigma}$ is neglected,
errors in $\tilde{\delta}$ and $\tilde{\varphi}$  
increase with the spatial length 
scale of perturbations decreasing and grow over time (see fig.~11 in Ref.~\cite{February:2013qza}).
In the present case, we see from Eqs.~\eqref{3:second2} and \eqref{3:second4} 
that the effects of $\tilde{\chi}^{(2)}$ and $\tilde{\varsigma}^{(2)}$ on $\tilde{\varphi}^{(2)}$ and $\tilde{\delta}^{(2)}$ 
may also increase with the spatial scale of perturbations decreasing, 
since the dominant terms related to $\tilde{\chi}^{(2)}$ and $\tilde{\varsigma}^{(2)}$ 
in the equations for $\tilde{\varphi}^{(2)}$ and $\tilde{\delta}^{(2)}$ are their spatial derivatives.
Thus, regarding the spatial scale dependence of the errors, our results seem to 
be consistent with those of FLCP.
As for the time dependence of the errors, our results imply
that the errors decrease over time, and this consequence contradicts the results obtained by FLCP.
We guess that the difference comes from the nonlinear effects of the spherical void;   
we focused on the situation in which the effects of a void structure are so small that they can be treated 
perturbatively, i.e., we assumed $\kappa\ll 1$, whereas FLCP assumed a highly nonlinear void structure 
at the present time (see Eq.~(10) in Ref.~\cite{February:2013qza}), i.e., the background LTB spacetime 
satisfies $\kappa \gg 1$ at the present time.

\section*{Acknowledgments}
KN was supported in part by JSPS Grant-in-Aid for Scientifc Research (C) (No. 25400265).



\end{document}